\documentstyle[eqsecnum,multicol,aps,prl,epsf]{revtex}

\date{\today}
\draft
\tighten
\begin{document}
\def\sqr#1#2{{\vcenter{\hrule height.3pt
      \hbox{\vrule width.3pt height#2pt  \kern#1pt
         \vrule width.3pt}  \hrule height.3pt}}}
\def\square{\mathchoice{\sqr67\,}{\sqr67\,}\sqr{3}{3.5}\sqr{3}{3.5}}
\def\today{\ifcase\month\or
  January\or February\or March\or April\or May\or June\or July\or
  August\or September\or October\or November\or December\fi
  \space\number\day, \number\year}

\def\Bbb{\bf}

\input amssym.tex

%%%%%%%%%%%%%%%%%%%%%%%%%%%%%%%%%%%%%%%%%%%%%%%%%%%%%%%%%%%%%
%%%%%%%%%%%%%%%%%%%%%%%%%%%%%%%%%%%%%%%%%%%%%%%%%%%%%%%%%%%%
%%%%%%%%%%%%%%%%%%%%%%%%%%%%%%%%%%%%%%%%%%%%%%%%%%%%%%%%%%%%%%%%%%
%%%%  MACROS:
%**********************************************************************
%GREEK LETTERS
\renewcommand{\a}{\alpha}
\renewcommand{\b}{\beta}
\newcommand{\g}{\gamma}           \newcommand{\G}{\Gamma}
\renewcommand{\d}{\delta}         \newcommand{\D}{\Delta}
\newcommand{\ve}{\varepsilon}
\newcommand{\eps}{\epsilon}
\newcommand{\k}{\kappa}
\newcommand{\ld}{\lambda}        \newcommand{\LD}{\Lambda}
\newcommand{\om}{\omega}         \newcommand{\OM}{\Omega}
\newcommand{\p}{\psi}             \newcommand{\PS}{\Psi}
\newcommand{\ro}{\rho}
\newcommand{\s}{\sigma}           \renewcommand{\S}{\Sigma}
\newcommand{\th}{\theta}         \newcommand{\T}{\Theta}
\newcommand{\f}{{\phi}}           \newcommand{\F}{{\Phi}}
\newcommand{\vf}{{\varphi}}
\newcommand{\y}{{\upsilon}}       \newcommand{\Y}{{\Upsilon}}
\newcommand{\z}{\zeta}
\newcommand{\X}{\Xi}
%************************************************************************
%  CAL. LETTERS
\newcommand{\cA}{{\cal A}}
\newcommand{\cB}{{\cal B}}
\newcommand{\cC}{{\cal C}}
\newcommand{\cD}{{\cal D}}
\newcommand{\cE}{{\cal E}}
\newcommand{\cF}{{\cal F}}
\newcommand{\cG}{{\cal G}}
\newcommand{\cH}{{\cal H}}
\newcommand{\cI}{{\cal I}}
\newcommand{\cJ}{{\cal J}}
\newcommand{\cK}{{\cal K}}
\newcommand{\cL}{{\cal L}}
\newcommand{\cM}{{\cal M}}
\newcommand{\cN}{{\cal N}}
\newcommand{\cO}{{\cal O}}
\newcommand{\cP}{{\cal P}}
\newcommand{\cQ}{{\cal Q}}
\newcommand{\cS}{{\cal S}}
\newcommand{\cR}{{\cal R}}
\newcommand{\cT}{{\cal T}}
\newcommand{\cU}{{\cal U}}
\newcommand{\cV}{{\cal V}}
\newcommand{\cW}{{\cal W}}
\newcommand{\cX}{{\cal X}}
\newcommand{\cY}{{\cal Y}}
\newcommand{\cZ}{{\cal Z}}
%***************************************************************
% CAPITAL LETTERS WITH HAT
\newcommand{\hA}{{\widehat A}}
\newcommand{\hB}{{\widehat B}}
\newcommand{\hC}{{\widehat C}}
\newcommand{\hD}{{\widehat D}}
\newcommand{\hE}{{\widehat E}}
\newcommand{\hF}{{\widehat F}}
\newcommand{\hG}{{\widehat G}}
\newcommand{\hH}{{\widehat H}}
\newcommand{\hI}{{\widehat I}}
\newcommand{\hJ}{{\widehat J}}
\newcommand{\hK}{{\widehat K}}
\newcommand{\hL}{{\widehat L}}
\newcommand{\hM}{{\widehat M}}
\newcommand{\hN}{{\widehat N}}
\newcommand{\hO}{{\widehat O}}
\newcommand{\hP}{{\widehat P}}
\newcommand{\hQ}{{\widehat Q}}
\newcommand{\hS}{{\widehat S}}
\newcommand{\hR}{{\widehat R}}
\newcommand{\hT}{{\widehat T}}
\newcommand{\hU}{{\widehat U}}
\newcommand{\hV}{{\widehat V}}
\newcommand{\hW}{{\widehat W}}
\newcommand{\hX}{{\widehat X}}
\newcommand{\hY}{{\widehat Y}}
\newcommand{\hZ}{{\widehat Z}}
%**********************************************************************
% LETTERS WITH HAT
\newcommand{\Ha}{{\widehat a}}
\newcommand{\Hb}{{\widehat b}}
\newcommand{\Hc}{{\widehat c}}
\newcommand{\Hd}{{\widehat d}}
\newcommand{\He}{{\widehat e}}
\newcommand{\Hf}{{\widehat f}}
\newcommand{\Hg}{{\widehat g}}
\newcommand{\Hh}{{\widehat h}}
\newcommand{\Hi}{{\widehat i}}
\newcommand{\Hj}{{\widehat j}}
\newcommand{\Hk}{{\widehat k}}
\newcommand{\Hl}{{\widehat l}}
\newcommand{\Hm}{{\widehat m}}
\newcommand{\Hn}{{\widehat n}}
\newcommand{\Ho}{{\widehat o}}
\newcommand{\Hp}{{\widehat p}}
\newcommand{\Hq}{{\widehat q}}
\newcommand{\Hs}{{\widehat s}}
\newcommand{\Hr}{{\widehat r}}
\newcommand{\Ht}{{\widehat t}}
\newcommand{\Hu}{{\widehat u}}
\newcommand{\Hv}{{\widehat v}}
\newcommand{\Hw}{{\widehat w}}
\newcommand{\Hx}{{\widehat x}}
\newcommand{\Hy}{{\widehat y}}
\newcommand{\Hz}{{\widehat z}}
%%%%%%%%%%%%%%%%%%%%%%%%%%%%%%%%%%%%%%%%%%%%%%%%%%%%%
%%%%%%%%%%%%%%%%%%%%%%%%%%%%%%%%%%%%%%%%%%%%%%%%%%%%%%%
%SPECIAL CHARACTERS
\newcommand{\deff}{\,\stackrel{\rm def}{\equiv}\,}
\newcommand{\lra}{\longrightarrow}
\newcommand{\ra}{\,\rightarrow\,}
\def\limar#1#2{\,\raise0.3ex\hbox{$\longrightarrow$\kern-1.5em\raise-1.1ex
\hbox{$\scriptstyle{#1\rightarrow #2}$}}\,}
\def\limarr#1#2{\,\raise0.3ex\hbox{$\longrightarrow$\kern-1.5em\raise-1.3ex
\hbox{$\scriptstyle{#1\rightarrow #2}$}}\,}
\def\limlar#1#2{\ \raise0.3ex
\hbox{$-\hspace{-0.5em}-\hspace{-0.5em}-\hspace{-0.5em}
\longrightarrow$\kern-2.7em\raise-1.1ex
\hbox{$\scriptstyle{#1\rightarrow #2}$}}\ \ }
\newcommand{\limm}[2]{\lim_{\stackrel{\scriptstyle #1}{\scriptstyle #2}}}
\newcommand{\wt}{\widetilde}
\newcommand{\os}{{\otimes}}
\newcommand{\da}{{\dagger}}
\newcommand{\stimes}{\times\hspace{-1.1 em}\supset}
\def\h{\hbar}
\newcommand{\ih}{\frac{\i}{\h}}
\newcommand{\exx}[1]{\exp\left\{ {#1}\right\}}
\newcommand{\ord}[1]{\mbox{\boldmath{$\cO$}}\left({#1}\right)}
%%%%%%%%%%%%%%%%%%
%% OPERATOR UNITY (modified 1)
\newcommand{\one}{{\leavevmode{\rm 1\mkern -5.4mu I}}}
%%%%%%%%%%%%%%%%%%
%% SETS OF NUMBERS
%    modified  Z                meaning:  integers
\newcommand{\Z}{Z\!\!\!Z}
%
%%%%%%%%%%%%%%%%%%
% The auxiliary definitions
\newcommand{\Ibb}[1]{ {\rm I\ifmmode\mkern
            -3.6mu\else\kern -.2em\fi#1}}
\newcommand{\ibb}[1]{\leavevmode\hbox{\kern.3em\vrule
     height 1.2ex depth -.3ex width .2pt\kern-.3em\rm#1}}
% are only used here:
%%%%%%%%%%%%%%%%%%%%%
%    modified   N               meaning:   natural numbers
\newcommand{\N}{{\Ibb N}}
%    modified   C               meaning:   complex numbers
\newcommand{\C}{{\ibb C}}
%    modified   R               meaning:   real numbers
\newcommand{\R}{{\Ibb R}}
%    modified   H               meaning:   quaternions
\newcommand{\HH}{{\Ibb H}}
\newcommand{\rational}{{\kern .1em {\raise .47ex
\hbox{$\scripscriptstyle |$}}
    \kern -.35em {\rm Q}}}
%%%%%%%%%%%%%%%%%%%%%%%%
%%%%%%%%%%%%%%%%%%%%%%%%%%%%%%%%%%%%%%%%
\newcommand{\bm}[1]{\mbox{\boldmath${#1}$}}
\newcommand{\intf}{\int_{-\infty}^{\infty}\,}
%%%%%% HILBERT SPACES %%%%%%%%%%%%%%%%
\newcommand{\LL}{\cL^2(\R^2)}
\newcommand{\LLS}{\cL^2(S)}
%%%%%%%%%%%%%%%%%%%%%%%%%%%%%%%%%%%%%%%
%%%%%% REAL AND IMAGINARY PARTS %%%%%%%%
\newcommand{\Ree}{{\cal R}\!e \,}
\newcommand{\Imm}{{\cal I}\!m \,}
%%%%%%%%%%%%%%%%%%%%%%%%%%%%%%%%%%%%%%%
%% SOME SPECIAL SYMBOLS IN ROMAN
\newcommand{\tr}{{\rm {Tr} \,}}
\newcommand{\er}{{\rm{e}}}
\renewcommand{\i}{{\rm{i}}}
\newcommand{\divv}{{\rm {div} \,}}
\newcommand{\id}{{\rm{id}\,}}
\newcommand{\ad}{{\rm{ad}\,}}
\newcommand{\Ad}{{\rm{Ad}\,}}
\newcommand{\const}{{\rm{\,const\,}}}
\newcommand{\rank}{{\rm{\,rank\,}}}
\newcommand{\diag}{{\rm{\,diag\,}}}
\newcommand{\sign}{{\rm{\,sign\,}}}
%%%%%%%%%%%%%%%%%%%%%%%%%%%%%%%%%%%%%%%%%%%%%%%%%%%%%%%%
%%%DERIVATIVES%%%
\newcommand{\pa}{\partial}
\newcommand{\pad}[2]{{\frac{\partial #1}{\partial #2}}}
\newcommand{\padd}[2]{{\frac{\partial^2 #1}{\partial {#2}^2}}}
\newcommand{\paddd}[3]{{\frac{\partial^2 #1}{\partial {#2}\partial {#3}}}}
\newcommand{\der}[2]{{\frac{{\rm d} #1}{{\rm d} #2}}}
\newcommand{\derr}[2]{{\frac{{\rm d}^2 #1}{{\rm d} {#2}^2}}}
\newcommand{\fud}[2]{{\frac{\delta #1}{\delta #2}}}
\newcommand{\fudd}[2]{{\frac{\d^2 #1}{\d {#2}^2}}}
\newcommand{\fuddd}[3]{{\frac{\d^2 #1}{\d {#2}\d {#3}}}}
\newcommand{\dpad}[2]{{\displaystyle{\frac{\partial #1}{\partial #2}}}}
\newcommand{\dfud}[2]{{\displaystyle{\frac{\delta #1}{\delta #2}}}}
\newcommand{\dd}{\partial^{(\ve)}}
\newcommand{\ddd}{\bar{\partial}^{(\ve)}}
\newcommand{\dfrac}[2]{{\displaystyle{\frac{#1}{#2}}}}
\newcommand{\dsum}[2]{\displaystyle{\sum_{#1}^{#2}}}
\newcommand{\dint}{\displaystyle{\int}}
\newcommand{\dg}{\!\not\!\partial}
\newcommand{\vg}[1]{\!\not\!#1}
%%%%%%%%%%%%%%%%%%%%%%%%%%%%%%%%%%%%%%%%%
%%%%%%%%%%%%%%%%%%%%%%%%%%%%%%%%%%%%%%%%%%%
%%%% BRA and KET                   %%%%%%%
\def\<{\langle}
\def\>{\rangle}
\def\lgl{\langle\langle}
\def\rgr{\rangle\rangle}
\newcommand{\bra}[1]{\left\langle {#1}\right|}
\newcommand{\ket}[1]{\left| {#1}\right\rangle}
\newcommand{\vev}[1]{\left\langle {#1}\right\rangle}
%%%%%%%%%%%%%%%%%%%%%%%%%%%%%%%%%%%%%%%%%%%%
%%%%%    EQUATIONS          %%%%%%%%%%%%%%
\newcommand{\be}{\begin{equation}}
\newcommand{\ee}{\end{equation}}
\newcommand{\bn}{\begin{eqnarray}}
\newcommand{\en}{\end{eqnarray}}
\newcommand{\bnn}{\begin{eqnarray*}}
\newcommand{\enn}{\end{eqnarray*}}
\newcommand{\ba}{\begin{array}}
\newcommand{\ea}{\end{array}}
\newcommand{\e}{\label}
\newcommand{\nbr}{\nonumber\\[2mm]}
\newcommand{\r}[1]{(\ref{#1})}
\newcommand{\refp}[1]{\ref{#1}, page~\pageref{#1}}
\renewcommand {\theequation}{\thesection.\arabic{equation}}
\renewcommand {\thefootnote}{\fnsymbol{footnote}}
%%%%%%%%%%%%%%%%%%%%%%%%%%%%%%%%%%%%%%%%
%%% SPACES      %%%%%%%%%%%%%%%%%%%%%%%%%%%
\newcommand{\qq}{\qquad}
\newcommand{\qqq}{\quad\quad}
%%%%%%%%%%%%%%%%%%%%%%%%%%%%%%%%%%%%%%%%%%%%%%%%
%%%%%%%    ADDITIONAL COMMANDS   %%%%%%%%%%%%%
\newcommand{\biz}{\begin{itemize}}
\newcommand{\eiz}{\end{itemize}}
\newcommand{\ben}{\begin{enumerate}}
\newcommand{\een}{\end{enumerate}}
%%%%%%%%%%%%%%%%%%%%%%%%%%%%%%%%%%%%%%%%%%%%%%%%%%%%%%%%%%%%%%%
\def\nc{noncommutative }
\def\ncy{noncommutativity }
\def\com{commutative }
%%%%%%%%%%%%%%%%%%%%%%%%%%%%%%%%%%%%%%%%%%%%%%%%%%%%%%%%%%%%%%

\title{Spin-Statistics and CPT Theorems\\
in Noncommutative Field Theory}

\author{ M. Chaichian, K. Nishijima$^{\dagger}$
\ \ and \ \ A. Tureanu}

\address{High Energy Physics Division, Department of
Physical Sciences,
University of Helsinki\\
\ \ {and}\\
\ \ Helsinki Institute of Physics,
P.O. Box 64, FIN-00014 Helsinki, Finland\\
$^{\dagger}$ Nishina Memorial Foundation\\
2-28-45 Honkomagome, Bunkyo-ku, Tokyo  113-8941, Japan}

\maketitle
\setcounter{footnote}{0}

\begin{abstract} We show that Pauli's spin-statistics relation remains valid in noncommutative quantum field theories (NC QFT), with the 
exception of some peculiar cases of noncommutativity between space and time. We also prove that, while the individual symmetries C and T, and in some cases
also P, are broken, the CPT theorem still holds {\it in general} for \nc field theories, in spite
of the inherent nonlocality and violation of Lorentz invariance.

\end{abstract}

\pacs{PACS: 11.30.Er, 02.04.Gh.
\hspace {3cm} HIP-2002-36/TH}
\vspace*{0.1cm}
\begin{multicols} {2}

\section{Introduction}
\setcounter{equation}{0}

Pauli's exclusion principle \cite{{Pauli1},{Pauli2}} and a more general formulation of it known as Pauli's spin-statistics relation or
theorem \cite{{Pauli3},{Pauli4}} is one of the most fundamental and important results in physics. This relation is responsible for the
entire structure of the matter and for its stability. Experimentally, the relation has been verified to high accuracy \cite{Capri}.
Theoretically up to now there has been no compelling argument or logical motivation for its breaking.

Within the framework of relativistic quantum field theory, Pauli demonstrated \cite{{Pauli3},{Pauli4}} the connection between spin and statistics,
based on the following requirements:

i) The vacuum is the state of lowest energy;

ii) Physical quantities (observables) commute with each other in two space-time points with a space-like distance;

iii) The metric in the physical Hilbert space is positive definite.

This well-known and celebrated result asserts that half-integer-spin fields (fermions), connected with the exclusion principle, can be
consistently quantized in accordance with Fermi-Dirac statistics, i.e. using anticommutation relations, while integer-spin fields (bosons) can be
consistently quantized in accordance with Bose-Einstein statistics, through commutation relations. Thus, the theorem is of wide applicability
provided we deal only with Fermi or Bose quantizations. A proof of the field commutation relations without reference to the specific form
of the interaction has been provided within the axiomatic formulation of quantum field theory \cite{Streater}.

A possible breaking of the spin-statistics relation in quantum field theory, due to a space-time noncommutativity, 
was previously suggested \cite{ChaichianSS}. In this letter, we show that such a violation could occur {\it only if} the space and time 
coordinates do not commute. 

At the same time we present a general proof that the CPT theorem remains valid in NC field theories, for general form of noncommutativity, although the
individual symmetries C,T and P are broken.

\section {Noncommutative quantum field theory and spin-statistics theorem}
\setcounter{equation}{0}

It is generally believed that the notion of space-time as a continuous manifold should break down at very short distances of the order of
the Planck length $\lambda_P\approx 1.6\times10^{-33}cm$. This would arise, e.g. from the process of measurement of space-time points
based on quantum mechanics and gravity arguments \cite{Dopli}. In measuring the points to great accuracy, one would need higher and higher
energy-momentum densities, which would finally create a black hole around the point, thus forbidding an infinite accuracy measurement.
Arguments for noncommutativity arise also from string theory with a constant antisymmetric background field, whose low-energy limit,
in some cases, turns up to be a NC QFT \cite{SW}. 
This in turn implies that our classical geometrical concepts may not be well
suited for the description of physical phenomena at very small distances. One such direction is to try to formulate physics on some \nc
space-time \cite{Dopli}-\cite{Connes}. If the concepts of \nc geometry are used, the notion of point as elementary geometrical entity is lost
and one may expect that an ultraviolet cutt-off appears \cite {Snyder} (see also \cite{Filk} where this expectation is shown not to occur
in general).

In a \nc space-time the coordinate operators satisfy the commutation relation:
\be\label{cr}
[x^{\mu},x^{\nu}]=\i\theta^{\mu\nu}\ , 
\ee 
where $\theta^{\mu\nu}$ is a general antisymmetric tensor of dimension 
(length)$^2$. In quantum field theory the operator character of the space-time coordinates (\ref{cr}) requires that the product of any two
field operators, $\phi(x)\Phi(x)$, be replaced by their $\star$-product (star-product), or Weyl-Moyal product, $\phi(x)\star\Phi(x)$. In the case when
$\theta^{\mu\nu}$ is a constant antisymmetric "tensor", the $\star$-product compatible with the associativity of field products is given by:
\be\label{star}
\phi(x)\star\Phi(x)=e^{\frac{i}{2}\theta^{\mu\nu}
\frac{\partial}{\partial{x^{\mu}}}\frac{\partial}{\partial{y^{\nu}}}}\phi(x)\Phi(y)\Big|_{x=y}.
\ee

In this case the physical quantities (observables) which are in general products of several field operators, are no more local quantities
and could therefore fail to fulfil the above requirement {\it ii)} (Sect. I) for the spin-statistics theorem to hold. For instance,
taking the product $:\phi^2(x):$ for a real scalar field with mass $m$, its \nc version $:\phi(x)\star\phi(x):$ could give a nonvanishing
equal-time commutation relation (ETCR):

\be\label{ETCR}
[:\phi(x)\star\phi(x):\ ,\ :\phi(y)\star\phi(y):]\Big|_{x_0=y_0}\neq 0,
\ee
where $:\ :$ denotes the normal ordering. In particular, while the vacuum expectation value of the ETCR (\ref{ETCR}) is still zero,
the matrix element between vacuum and a two-particle state, on a $d$-dimensional space, when Bose statistics is used, is:
\begin{eqnarray}\label{matr_el}
&\langle&0|[:\phi(x)\star\phi(x):\ ,\ :\phi(y)\star\phi(y):]\Big|_{x_0=y_0}|p,p'\rangle\cr &=&
-\frac{2i}{(2\pi)^{2d}}\frac{1}{\sqrt{\omega_p\omega_{p'}}}(e^{-ip'x-ipy}+e^{-ipx-ip'y})\cr
&\int&\frac{d\vec{k}}{\omega_k}\sin[\vec{k}(\vec{x}-\vec{y})]
\cos(\frac{1}{2} \theta^{\mu\nu}k_{\mu}p_{\nu})\cos(\frac{1}{2}\theta^{\mu\nu}k_{\mu}p'_{\nu}),
\end{eqnarray}
where $\omega_k=k_0=\sqrt{{\vec k}^2+m^2}$ and $\vec{k}=(k_1,...,k_d)$.
The r.h.s. of (\ref{matr_el}) is nonzero only when $\theta^{0i}\neq0$. This statement holds for the matrix elements of ETCR of two
observables expressed as any power of bosonic fields $\phi(x)$, $\phi(y)$ and their derivatives, with
$\star$-product analogous to (\ref{matr_el}), and as well for products of spinor fields $\bar{\psi}(x)$, $\psi(y)$ and their derivatives, 
with
anti-commutation relation used in the latter case. It is known, however, that for such field theories with space-time
noncommutativity ($\theta^{0i}\neq 0$) there appears the violation of unitarity \cite{Gomis} as well as the violation of causality at the
macroscopic level, such as in scattering processes \cite {Seiberg}. Indeed, while the low-energy limit of string theory in a constant 
antisymmetric
background field $B^{mn}$, which exhibits noncommutativity, reduces to field theory with the $\star$-product when $\theta^{0i}=0$, for
the case $\theta^{0i}\neq 0$ there is no corresponding low-energy field theory limit. 

Still, there is the exception of the field
theories with light-like noncommutativity, $\theta^{\mu\nu}\theta_{\mu\nu}=0$, i.e. $\theta^{0i}=-\theta^{1i}$, 
for which unitarity is preserved \cite{Aharony}. In this case, however, the 
microcausality in
the sense of ETCR (\ref{matr_el}) is still violated. For instance, if we consider $p_\mu=p'_\mu$ ($\mu=0,1,2$) and $x-y\equiv z$, 
then the integral
in (\ref{matr_el}) becomes:
\be\label{int}
I=\frac{\pi}{4}\frac{\cos({m\sqrt{(\theta p_2)^2-(\theta p_2-|z_1|)^2-(\theta \omega_p-\theta p_1 -|z_2|)^2}})}
{\sqrt{(\theta p_2)^2-(\theta p_2-|z_1|)^2-(\theta \omega_p-\theta p_1 -|z_2|)^2}}\ ,
\ee
with $\theta\equiv\theta^{0i}$ (taken to be positive), for
\bn
0&<&\theta p_2-|z_1|<\theta p_2\ ,\cr
0&<&\theta \omega_p-\theta p_1-|z_2|<\sqrt{(\theta p_2)^2-(\theta p_2-|z_1|)^2}\ ,
\en
i.e. for the values
\bn\label{cond}
0&<&|z_1|<\theta p_2\ ,\cr
\theta (\omega_p- p_1- p_2)&<&|z_2|<\theta (\omega_p- p_1)\ . 
\en
%(if the conditions (\ref{cond}) are not fulfilled, the integral is $0$).
In this case (\ref{int}) is nonzero, allowing therefore the violation of the spin-statistics relation (see, however, the remarks in Sect.
IV).
Note that the conditions (\ref{cond}) are compatible with the "advanced displacements" responsible for the violation of macrocausality in
\cite{Seiberg}. Also, in the second condition of the set, the lower limit for $|z_2|$, which is positive and finite, shows the intrinsic
indetermination in the coordinate which does not commute with time, once we took $x_0-y_0\equiv z_0=0$.

If the field theory with light-like noncommutativity is indeed the low-energy limit of string theory, as stated in
\cite{Aharony}, it is then intriguing that the theory is unitary but acausal (as it is known that a low-energy effective theory 
should not necessarily be
unitary, as is the case, e.g., for the Fermi four-spinor interaction).

\section{CPT Theorem in NC Field Theories}

The CPT theorem \cite{Luders,{CPT}} (see also \cite{Streater} for a review) is of a universal nature in that it is valid in
all the known field theories. Here we shall recapitulate
essential features of the CPT transformation and then extend the CPT theorem to \nc field theories.

First, we shall summarize the common properties of anti-unitary transformations, including time reversal and CPT transformation. An
antiunitary transformation denoted hereafter by ${\blacklozenge}$ is a generalization of complex conjugation and satisfies
\be
(\Psi^{\blacklozenge}\ ,\Phi^{\blacklozenge})=(\Phi\ ,\Psi)\ .
\ee
The transformation of state vectors corresponds to the Schrödinger picture and we can also attribute the same transformation to operators
corresponding to the Heisenberg picture by
\be
(\Psi^{\blacklozenge}\ ,Q\Phi^{\blacklozenge})=(\Phi\ ,Q^{\blacklozenge}\Psi)\ .
\ee

In what follows we shall mainly discuss the latter approach.

a) The transformation of operators obeys the following rules:
\bn
(c_1A+c_2b)^{\blacklozenge}&=&c_1A^{\blacklozenge}+c_2B^{\blacklozenge}\ (linearity)\ ,\cr
(AB)^{\blacklozenge}&=&B^{\blacklozenge} A^{\blacklozenge},
\en
where $c_1$ and $c_2$ are $c$-number coefficients.

b) Let us assume that
\be
Q^{\blacklozenge}=\epsilon Q, (\epsilon=\pm 1)
\ee
and that $\Psi$ is an eigenstate of $Q$ with the eigenvalue $q$,
\be
Q\Psi=q\Psi;
\ee
then $\Psi^{\blacklozenge}$ is also an eigenstate of $Q$ and
\be
Q\Psi^{\blacklozenge}=\epsilon q\Psi^{\blacklozenge}.
\ee

\subsection{The CPT transformation of local elementary fields}

In what follows we shall use the symbol ${\blacklozenge}$ exclusively for the CPT transformation and we shall first define it for local elementary
fields. Let $\psi_\alpha$, $\bar{\psi}_\alpha$ and $\phi_{\lambda_1...\lambda_n}$ be local {\it elementary} fields representing spinors and
tensors, respectively; then the CPT transformation is specified by \cite{Nishijima}:
\bn\label{CPTtr}
\psi_\alpha^{\blacklozenge}(x)&=&(i\gamma_5)_{\alpha\beta}\psi_\beta(-x)\ ,\cr
\bar{\psi}_\alpha^{\blacklozenge}(x)&=&\bar{\psi}_\beta(-x)(i\gamma_5)_{\beta\alpha}\ , \cr
\phi_{\lambda_1...\lambda_n}^{\blacklozenge}(x)&=&(-1)^n\phi_{\lambda_1...\lambda_n}(-x).
\en
This set of rules completely specifies the transformation of any local elementary field carrying definite spinor and/or tensor indices.
Then the CPT theorem for local field theories can be formulated in the following form:

\subsection{CPT theorem for local fields}

Let $\psi_\alpha$, $\bar{\psi}_\alpha$ and $\phi_{\lambda_1...\lambda_n}$ be local but {\it composite} fields representing spinors and
tensors, respectively; then they are transformed exactly in the same form as eq. (\ref{CPTtr}) for local elementary fields.

In what follows we shall clarify the significance of this theorem.

1) Let us consider local composite scalar fields of which free and interaction Lagrangian densities as well as interaction Hamiltonian
densities are typical members, and we have:
\be\label{L_transf}
{\cal L}_f^{\blacklozenge}(x)={\cal L}_f(-x)\ ,\ \ {\cal L}_{int}^{\blacklozenge}(x)={\cal L}_{int}(-x)\ ,
\ee
\be\label{theorem}
{\cal H}_{int}^{\blacklozenge}(x)={\cal H}_{int}(-x)\ .
\ee

In \cite{Nishijima}, eq. (\ref{theorem}) has been referred to as the CPT theorem and its proof has been given there so that we skip it.

When asymptotic conditions are valid, the CPT invariance of the $S$ matrix follows from it:
\be
S^{\blacklozenge}=S\ .
\ee 

2) Next, let $\Phi_\lambda$ be a local composite vector field and $\phi_\lambda$ a local elementary vector field, respectively; then a
composite scalar field $\Phi=\phi_\lambda\Phi_\lambda$ is transformed as (\ref{L_transf}) or (\ref{theorem}) and $\phi_\lambda$ as
(\ref{CPTtr}).
From the above information we deduce:
\be
\Phi_\lambda^{\blacklozenge}(x)=-\Phi_\lambda(-x),
\ee
and similarly we can prove eq. (\ref{CPTtr}) for spinors and tensors. As an example of local composite vector fields we choose the electric
current density $j_\lambda(x)$; then the conserved electric charge $Q$ transforms as:
\be
Q^{\blacklozenge}=\int d^3xj_0^{\blacklozenge}(x)=-\int d^3x j_0(-x)=-Q\ .
\ee
3) The energy-momentum vector $P_\lambda$ can be expressed as the space integral of the energy-momentum tensor of the second rank.
Therefore, we immediately conclude
\be
P_\lambda^{\blacklozenge}=P_\lambda\ .
\ee
4) The generators of the Lorentz transformation $M_{\rho\sigma}$ can be expressed as the space integral of a tensor of the third rank, so
that we have:
\be
M_{\rho\sigma}^{\blacklozenge}=-M_{\rho\sigma}\ .
\ee
This indicates that the spin of a particle defined in terms of the Pauli-Lubanski operator should reverse its direction under CPT.

In general, the CPT transformation of an operator is determined by the tensorial rank of its density.

5) We assume the validity of the LSZ asymptotic conditions \cite{LSZ}; then on the basis of their definition of the asymptotic fields it is
straightforward to show that the CPT transformation turns incoming fields into outgoing fields and vice versa.

\subsection{CPT theorem for \nc fields}

The validity of CPT theorem for \nc QED has been discussed in \cite{Shahin}, where it was concluded that CPT is {\it accidentally} preserved,
although the charge conjugation and time reversal symmetries are broken due to noncommutativity. However, in \cite{Shahin} the
specific version of NC QED of \cite{Haya} was studied, where the photon couples only to particles with the electric charges $+1, -1$ and
$0$. The latter is usually referred to as the "charge quantization problem".

In the following, we shall show the {\it general} validity of the CPT heorem for any \nc quantum field theory of the type described in
Sect. II.

Let ${\cal H}(x)$ be the Weyl-Moyal product (\ref{star}) of field operators representing the interaction Hamiltonian in a \nc field theory. It is understood
that ${\cal H}(x)$ stands for a normal product in the interaction representation. The CPT theorem is given by
\be
{\cal H}_{int}^{\blacklozenge}(x)={\cal H}_{int}(-x)\ .
\ee
In order to prove it we shall choose as an illustration a $n$-linear form for ${\cal H}(x)$, namely,
\bn\label{Hamiltonian}
&{\cal H}&(x)=\cr
&=&\sum_{i_1...i_n}f_{i_1...i_n}\phi^1_{i_1}(x)\star ...\star \phi^n_{i_n}(x)\cr
&=&e^D\sum_{i_1...i_n}f_{i_1...i_n}\phi^1_{i_1}(x_1) ...\phi^n_{i_n}(x_n)
|_{x_1=...=x_n\equiv x}
\ ,
\en
where $i_j$ with $j=1,..., n$ stand for spinorial or tensorial indices and the coefficients $f_{i_1...i_n}$ are so chosen as to make ${\cal H}(x)$ a
scalar under proper Lorentz transformations, {\it in the local limit}. $D$ stands for the differential operator of the form
\be\label{phase}
D=e^{\frac{i}{2}\theta^{\mu\nu}(\partial_\mu^{x_1}\partial_\nu^{x_2}+\partial_\mu^{x_2}\partial_\nu^{x_3}
+...+\partial_\mu^{x_{n-1}}\partial_\nu^{x_n}})\ ,
\ee
with general $\theta^{\mu\nu}$.
Then the CPT transform of ${\cal H}(x)$ is given by:
\bn\label{H_tr}
&{\cal H}&^{\blacklozenge}(x)=\cr
&=&e^D\sum_{i_1...i_n}f_{i_1...i_n}\phi^{n\blacklozenge}_{i_n}(x_n)...\phi^{1\blacklozenge}_{i_1}(x_1)|_{x_1=...
=x_n\equiv x}\cr
&=&e^D\sum_{i_1...i_n}f'_{i_1...i_n}\phi^n_{i_n}(-x_n)...\phi^1_{i_1}(-x_1)|_{x_1=...
=x_n\equiv x}\ ,
\en
where $f'$ is given by
\be
f'_{i_1...i_n}=(-1)^{F/2}f_{i_1...i_n},
\ee
and $F$ stands for the number of the Fermi fields involved in ${\cal H}(x)$. When we reverse the order of multiplication back to the original
one in (\ref{Hamiltonian}), we obtain:
\bn
&{\cal H}&^{\blacklozenge}(x)=\cr
&=&e^D\sum_{i_1...i_n}f_{i_1...i_n}\phi^1_{i_1}(-x_1) ...\phi^n_{i_n}(-x_n)|_{x_1=...
=x_n\equiv x}\cr
&=&\sum_{i_1...i_n}f_{i_1...i_n}\phi^1_{i_1}(-x)\star ...\star \phi^n_{i_n}(-x)\cr 
&=&{\cal H}(-x)\ .
\en
Thus the CPT theorem is valid not only in local field theories but also in \nc field theories.

This can be also seen from the fact that, when we expand the interaction Hamiltonian density in powers of $\theta$,
the first term is the local limit of the Hamiltonian expressed in terms
of the Weyl-Moyal product. It is a local but composite scalar density. The
coefficients of other terms are local but composite tensor fields of
even ranks obtained by differentiating the fields involved in the first
term, an even number of times. Therefore, they transform in the same way as
the first term under CPT. From this point of view it is intuitively
clear that the Hamiltonian density expressed in terms of the Weyl-Moyal product
transforms in the same way as the local ones under CPT.

As seen from the proof presented above, the CPT theorem is valid for any form of noncommutativity (general $\theta^{\mu\nu}$).

{\it Individual discrete transformations P, C and T}

The individual transformations P, C and T are violated in many cases and we shall comment on them only by comparison with the local
(commutative) limit of the \nc field theory in question. In the case of only space-space noncommutativity ($\theta^{0i}=0$), the parity of a \nc field theory is the same as for
its commutative limit, while charge conjugation and time reversal are broken, even if they hold for the  commutative
limit. This is due to the fact that C and T imply a complex conjugation, that would change the sign of the phase in (\ref{phase}). In the
case of a space-time noncommutative theory ($\theta^{0i}\neq 0$) - whose commutative limit is P, C and T invariant - 
all these discrete transformations are violated, as in the NC QED case \cite{Shahin}.

\section{Conclusions}

In this letter we have studied the implications of noncommutativity of space and time on the validity of spin-statistics and CPT theorems. 
The violation of
Lorentz invariance as well as the intrinsic nonlocality of \nc field theories may suggest that a (presumably very small, of the order of 
$|\theta^{\mu\nu}|m^2$) breaking of these two fundamental theorems
might be possible. 

We have found that the CPT theorem is
{\it generally} valid in NC FT, irrespective of the form of the noncommutativity parameter $\theta^{\mu\nu}$ involved. 
As for the spin-statistics theorem, we have found that it holds
in the case of field theories with space-space noncommutativity, which can be obtained as a low-energy limit from the string theory. 

A violation of the spin-statistics relation due to the noncommutativity of space and time ($\theta^{0i}\neq0$) can not be justified, 
given the pathological character of
such theories. The case of light-like noncommutativity ($\theta^{\mu\nu}\theta_{\mu\nu}=0$), which is compatible with unitarity, deserves,
however more attention. 

In conclusion, it is of importance to study further the light-like case, as to determine whether it can indeed be obtained as a low-energy
limit of string theory. Questions concerning a possible breaking of the spin-statistics relation are of outmost importance, since such a 
violation, no matter how small, would have a crucial impact on the
structure and the stability of matter in the Universe. The issue, on the other hand, is of fundamental interest by itself, since up to now no
theoretical argument or motivation for such a breaking has been presented. The present work is a possible step in this direction.

\vskip 0.3cm
{\bf{Acknowledgements}} 
We are grateful to L. Alvarez-Gaum\'{e}, A. Kobakhidze, Y. Liao, C. Montonen, D. Polyakov and P. Pre\v{s}najder  
for useful discussions. 

The financial support of the Academy of Finland under the Project No. 54023
is acknowledged.

\end{multicols}

\begin{thebibliography}{99}

%%%%%%1 
\bibitem{Pauli1} W. Pauli, {\it Z. Physik} {\bf 31} (1925) 765
(reprinted in \cite{Pauli2}).

%%%%%%2
\bibitem{Pauli2}
W. Pauli, {\it Collected Scientific Papers}, vol. 2, Eds. R. Kronig and V. 
F. Weisskopf, J. Wiley \& Sons, New York, 1964.

%%%%%3
\bibitem{Pauli3}
W. Pauli, {\it Phys. Rev.} {\bf 58} (1940) 716.

%%%%%4
\bibitem{Pauli4}
W. Pauli, {\it Progr. Theor. Phys. (Kyoto)} {\bf 5} (1950) 526.

%%%%%5
\bibitem{Capri}
See e.g., {\it Proceedings of the Conference on Spin-Statistics Connection 
and Commutation Relations}, Capri, 2000, Eds. R. C. Hilborn and G. M. 
Tino, AIP Conference Proceedings, vol. 545, 2000.

%%%%%%6
\bibitem{Streater}
G. F. Streater and A. S. Wightman, {\it PCT, Spin, Statistics and All 
That}, W. A. Benjamin, Inc., New York, 1964, and references therein.


%%%%%7
\bibitem{ChaichianSS}
M. Chaichian, Talk presented in \cite{Capri}, p. 129.


%%%%%%9
\bibitem{Dopli}
S. Doplicher, K. Fredenhagen and J. E. Roberts, {\it Phys. Lett} {\bf 
B331} (1994) 39; {\it Comm. Math. Phys.} {\bf 172} (1995) 187.

%%%%%8
\bibitem{SW}
N. Seiberg and E. Witten, {\it JHEP} {\bf 9909} (1999) 32, hep-th/9908142 
and references therein.

%%%%%%10
\bibitem{Snyder}
H. S. Snyder, {\it Phys. Rev} {\bf 71} (1947) 38.
According to a survey by J. Wess, Heisenberg conveyed to R. Peierls his 
idea that noncommutating space coordinates could resolve the problem of 
infinite self-energies. Apparently, Peierls also described it to Pauli \cite{Jackiw}. It 
is interesting that it was Pauli who pursued further this idea by 
explaining it to Oppenheimer who told it to Snyder, what could have brought to a possible violation of his own spin-statistics theorem.

%%%%%11
\bibitem{Connes}
A. Connes, {\it Noncommutative Geometry}, Academic Press, New York, 1994.

%%%%%12
\bibitem{Filk}
T. Filk, {\it Phys. Lett.} {\bf B 376} (1996) 53;

M. Chaichian, A. Demichev and P. Pre\v{s}najder, {\it Nucl. Phys.} {\bf B 567} 
(2000) 360, hep-th/9812180.

%%%%%18
\bibitem{Jackiw}
R. Jackiw, {\it Nucl. Phys. Proc. Suppl.} {\bf 108} (2002) 30, hep-th/0110057.

%%%%%%13
\bibitem{Gomis}
J. Gomis and T. Mehen, {\it Nucl. Phys.} {\bf B 591} 
(2000) 265, hep-th/0005129.

%%%%%%15
\bibitem{Seiberg}
N. Seiberg, L. Susskind and N. Toumbas, {\it JHEP} {\bf 0006} (2000) 044, hep-th/0005015.

%%%%%%
\bibitem{Aharony}
O. Aharony, J. Gomis and T. Mehen, {\it JHEP} {\bf 0009} (2000) 023, hep-th/0006236.

%%%%%%%%%%%
\bibitem{Luders}
G. L\"{u}ders, {\it Dansk. Mat. Fys. Medd.} {\bf 28} (1954) 5;


%%%%%%%%%%%%%
\bibitem{CPT}
W. Pauli, {\it Niels Bohr and the Development of Physics}, W. Pauli (ed.), Pergamon Press, New York, 1955.

%%%%%%%%
\bibitem{Nishijima}
K. Nishijima, {\it Fundamental Particles}, W. A. Benjamin, Inc., New York, 1963.

%%%%%%%%
\bibitem{LSZ}
H. Lehmann, K. Symanzik and W. Zimmermann, {\it Nuovo Cimento} {\bf 1} (1955) 1425; {\bf 6} (1957) 319.

%%%%%%%%%
\bibitem{Shahin}
M. M. Sheikh-Jabbari, {\it Phys. Rev. Lett.} {\bf 84} (2000) 5265, hep-th/0001167.

%%%%%%%%
\bibitem{Haya}
M. Hayakawa, {\it Phys. Lett.} {\bf B 478} (2000) 394, hep-th/9912094.

\end{thebibliography}
\end{document}